\documentclass[aps,twocolumn,showpacs,amsfonts,amsmath,superscriptaddress,amssymb,prl]{revtex4}

\usepackage{graphicx}% Include figure files
\usepackage{dcolumn}% Align table columns on decimal point
\usepackage{bm}% bold math
\usepackage{longtable}

\begin{document}

\title{Electronic reconstruction at $\rm SrMnO_{3}-LaMnO_{3}$ superlattice
interfaces}

\author{\c{S}erban Smadici}
  \affiliation{Frederick Seitz Materials Research Laboratory, University of Illinois, Urbana, Illinois 61801, USA}%

\author{Peter Abbamonte}
  \affiliation{Frederick Seitz Materials Research Laboratory, University of Illinois, Urbana, Illinois 61801, USA}%

\author{Anand Bhattacharya}
  \affiliation{Center for Nanoscale Materials and Materials Science Division, Argonne National Laboratory, Illinois 60439, USA}%

\author{Xiaofang Zhai}
  \affiliation{Frederick Seitz Materials Research Laboratory, University of Illinois, Urbana, Illinois 61801, USA}%

\author{Andrivo Rusydi}
  \affiliation{Institute of Applied Physics, University of Hamburg, D-20355, Germany}%

\author{James N. Eckstein}
  \affiliation{Frederick Seitz Materials Research Laboratory, University of Illinois, Urbana, Illinois 61801, USA}%

\author{Samuel D. Bader}
  \affiliation{Center for Nanoscale Materials and Materials Science Division, Argonne National Laboratory, Illinois 60439, USA}%

\author{Jian-Min Zuo}
  \affiliation{Frederick Seitz Materials Research Laboratory, University of Illinois, Urbana, Illinois 61801, USA}%

\begin{abstract}
We use resonant soft x-ray scattering to study electronic
reconstruction at the interface between the Mott insulator
$\rm~LaMnO_{3}$ and the ``band" insulator $\rm~SrMnO_{3}$.
Superlattices of these two insulators were shown previously to have
both ferromagnetism and metallic tendencies [Koida \emph{et al.},
Phys. Rev. B \textbf{66}, 144418 (2002)]. By studying a judiciously
chosen superlattice reflection we show that the interface density of
states exhibits a pronounced peak at the Fermi level, similar to
that predicted by Okamoto \emph{et al.} [Phys. Rev. B \textbf{70},
241104(R) (2004)]. The intensity of this peak correlates with the
conductivity and magnetization, suggesting it is the driver of
metallic behavior. Our study demonstrates a general strategy for
using RSXS to probe the electronic properties of heterostructure
interfaces.
\end{abstract}

\pacs{71.45.Gm, 74.25.Jb, 78.70.Ck}

\maketitle

The interface between two correlated electron systems may have
properties that are qualitatively different from either of the two
constituents, providing a potential route to new devices and
physical properties.~\cite{ABV2006} An example is the interface
between the $d^{1}$ Mott insulator $\rm~LaTiO_{3}$ (LTO) and the
$d^{0}$ band insulator $\rm~SrTiO_{3}$ (STO), which experiments on
superlattice heterostructures have suggested is
metallic~\cite{OMGH2002}. Dynamical mean field theory (DMFT) studies
of this interface have suggested that the metallic behavior is
driven by interfacial electronic reconstruction, characterized by
the appearance of a quasiparticle peak at the Fermi level in the
density of states of the interface layer~\cite{OM2004,OM2005}.
Efforts to find this peak with angle-integrated photoemission
studies of LTO-STO superlattices observed a small Fermi surface
crossing~\cite{TWTH2006}. However, a peak in the density of states
clearly associated with the buried interface has not yet been
observed.

Another example of a Mott insulator - band insulator interface is
$\rm~LaMnO_{3}-SrMnO_{3}$.  $\rm~LaMnO_{3}$ (LMO) is a Mott
insulator with a $t_{2g}^{3}e_{g}^{1}$ configuration while
$\rm~SrMnO_{3}$ (SMO), which has a $t_{2g}^{3}e_{g}^{0}$
configuration, can be considered a high-spin band insulator because
its $e_{g}$ shell is empty. Therefore, apart from the presence of
the core $t_{2g}$ spins, LMO-SMO is analogous to the LTO-STO system.
A recent DMFT calculation predicted that the LMO-SMO interface
should be a ferromagnetic metal, governed by double-exchange hopping
of the $e_{g}$ electrons.~\cite{LOM2006} Large period LMO-SMO
heterostructures synthesized previously were shown to have both
ferromagnetism and metallic tendencies~\cite{KLFI2002}, however it
is not clear whether this derives from electronic reconstruction at
the interfaces.

In this Letter we present a study of LMO-SMO superlattices with
resonant soft x-ray scattering (RSXS). By judiciously choosing the
thicknesses of the LMO and SMO sublayers, we obtained a structure
whose third order superlattice reflection is directly sensitive to
the density of states (DOS) of the interface $\rm MnO_{2}$ layer. We
show that the DOS of the interface layer exhibits a pronounced
quasiparticle resonance at $E_{F}$ whose intensity correlates with
the magnetization and conductivity of the overall structure. Our
results confirm the predictions of Ref.~\cite{OM2004,OM2005} and
demonstrate a general strategy for using RSXS to study the
electronic structure of heterostructure interfaces.

\begin{figure}
\centering\rotatebox{0}{\includegraphics[scale=0.45]{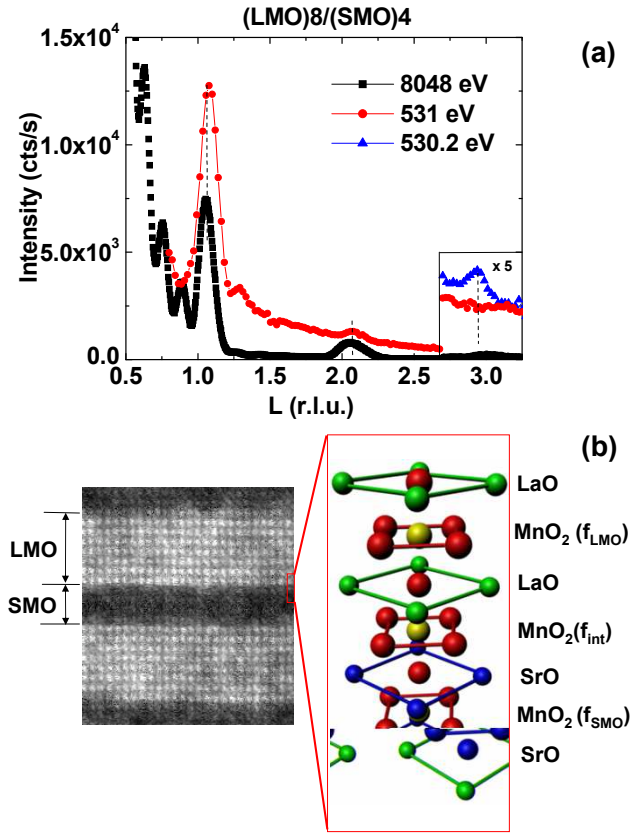}}
\caption{\label{fig:Figure2} (a) Specular x-ray scattering from the
$n=4$ superlattice with non-resonant hard x-rays (black circles),
non-resonant soft x-rays (red circles), and soft x-rays tuned close
to the Fermi energy (blue circles). Data are plotted on a linear
scale to emphasize height differences between the peaks. (b) (left)
STEM image of the $n=4$ structure and (right) a drawing showing the
interface region with scattering factors defined for the various
atomic planes.}
\end{figure}

Superlattices consisting of seven periods of (8 $\times$ LMO + 4
$\times$ SMO) and six periods of (10 $\times$ LMO + 5 $\times$ SMO)
layers were grown on $\rm~SrTiO_{3}$ (STO) substrates by molecular
beam epitaxy. The samples will be denoted (LMO)$2n$/(SMO)$n$ with
$n=4$ or $n=5$ in the following. In order to avoid oxygen vacancies,
which can cause anomalous metallic behavior through electron
doping~\cite{HBBC2007,MNOG2004}, the samples were both grown and
post-annealed in flowing ozone. Large amplitude RHEED oscillations
indicated two-dimensional epitaxial growth and STEM images showed
well-defined superlattice interfaces (Fig.~\ref{fig:Figure2}b). The
SMO and LMO overlayers in these structures are under $+2.8 \%$ and
$-2.1\%$ strain, respectively, which likely alters the exact pattern
of orbital and magnetic order in the
sublayers, but will not by itself induce
metallic behavior~\cite{WBNM2006,AM2001}. X-ray reflectivity
measurements showed clear interference fringes
(Fig.~\ref{fig:Figure2}a), indicating flat interfaces over
macroscopic distances.

RSXS measurements were carried out at the soft x-ray undulator
beamline X1B at the National Synchrotron Light Source in a 10-axis
UHV diffractometer~\cite{AVR2002,ARSG2005}. Measurements were done
in specular geometry, i.e. with the momentum transfer perpendicular
to the plane of the heterostructure.  In this article momenta will
be denoted in reciprocal units of the superlattice, i.e. Miller
index $L$ corresponds to a momentum $Q = 2 \pi L/d$, where $d$ is
the repeat period of the structure. The incident light was polarized
in the scattering plane ($\pi$ polarization) with the channeltron
detector integrating over both final polarizations, i.e. both the
$\pi \rightarrow \sigma$ and $\pi \rightarrow \pi$ scattering
channels. Bulk sensitive RSXS measurements were performed at the Mn
$\rm~L_{3,2}$ edges as well as the O K edge, which probes the $3d$
levels through hybridization, with the incident bandwidth set to
$\rm~0.2~eV$ resolution. X-ray absorption (XAS) measurements were
done \emph{in situ} in total electron yield (TEY) mode and probed
the top layer of the sample surface.

\begin{figure}
\centering\rotatebox{0} {\includegraphics[scale=0.45]{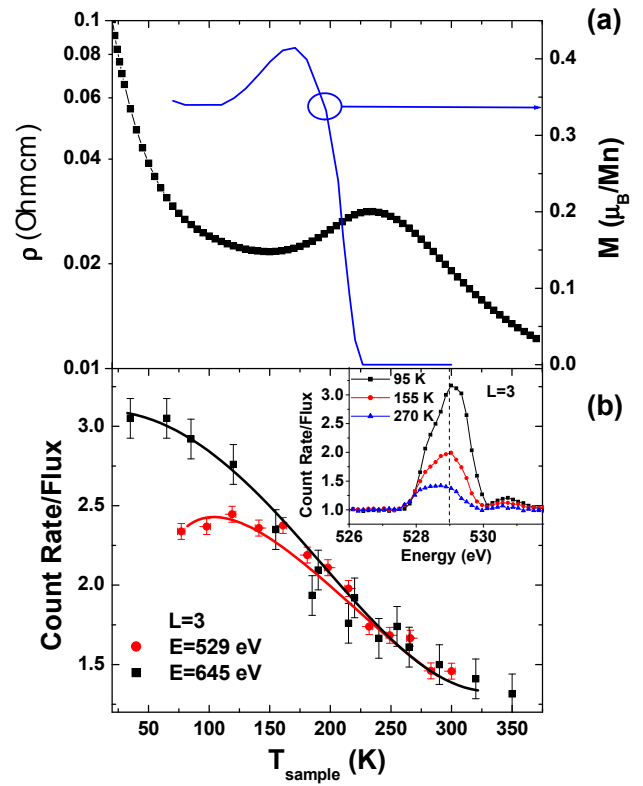}}
\caption{\label{fig:Figure1} (a). Temperature dependence of the
conductivity and magnetization of the $n=4$ structure when cooled in
$\rm 50~G$ magnetic field.~\cite{AKZ2007} (b). Temperature
dependence of the $L=3$ reflection at O K and $\rm~Mn~L_{3}$ edges
and $L=3$ for the $n=4$ sample. The inset shows temperature
dependent scans at O K edge. Two quasiparticle peaks at the O K
onset can be resolved.}
\end{figure}

\begin{figure}
\centering\rotatebox{0}{\includegraphics[scale=0.45]{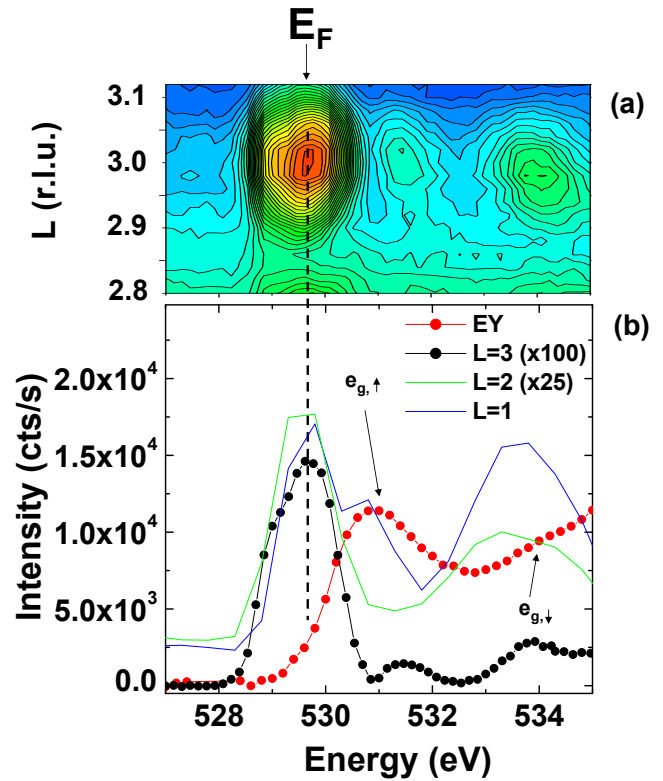}}
\caption{\label{fig:Figure3} (a). Resonance profile at $L=3$ and O K
edge for the $n=4$ sample. The strongest resonance is at the energy
of the doped holes and $L=3$. (b). Scattering at $L=3$ and
$T=90\rm~K$ (black), at $L=1$ (blue) and $L=2$ (green), compared to
XAS data (red). The strong resonant enhancement before the O K edge
is indicative of an electronic effect. The peaks at higher energy
are aligned with features in the absorption spectra, as assigned in
Ref. \cite{AGFF1992}.  The peaks at $529~\rm eV$ are interface
states split-off from bulk bands.}
\end{figure}

\begin{figure}
\centering\rotatebox{0} {\includegraphics[scale=0.45]{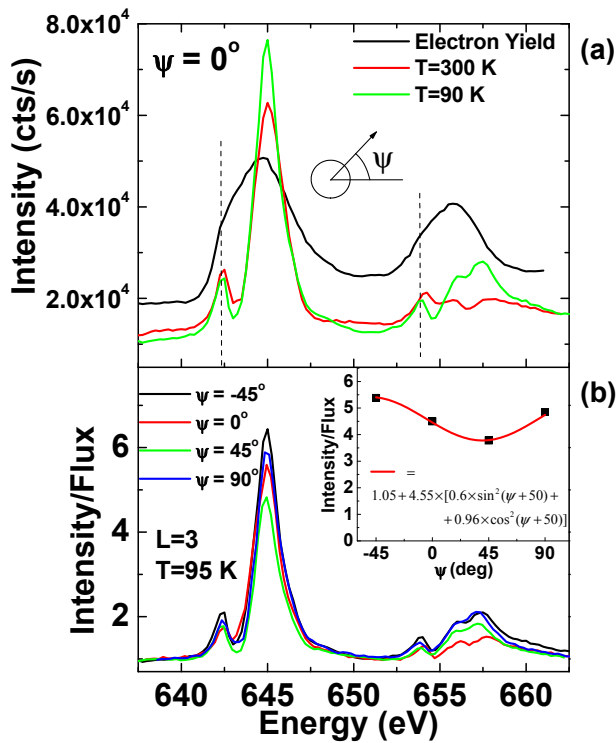}}
\caption{\label{fig:Figure4} (a). Energy profiles near
$\rm~Mn~L_{3,2}$ edges at $L=3$ for two temperatures (red and green)
and sample absorption measured by electron yield (black). In
contrast to the $L_{3}$ edge, the $L_{2}$ edge resonance profile is
strongly affected by the core hole. (b) $L=3$ scattering at $\rm
645~eV$ for a few different azimuthal orientations. The inset shows
the measured azimuthal dependence and a fit curve.}
\end{figure}

Prior to RSXS studies the superlattices were characterized with XAS,
resistivity, magnetization, hard x-ray diffraction, and scanning
transmission electron microscopy (STEM). XAS probes only the
near-surface region which in the current case is the top LMO layer.
The spectra (Figs.~\ref{fig:Figure3}b and~\ref{fig:Figure4}a; note
the location of the Fermi energy) closely resemble past studies of
LMO powders \cite{AGFF1992} indicating good surface quality. The
resistivity of both samples exhibits a crossover to metallic
behavior below $T_c \sim\rm~225~K$ (Fig.~\ref{fig:Figure1}a and
Ref.~\cite{AKZ2007}). At this temperature ferromagnetism also
appears, with an average saturation magnetic moment at low
temperatures of $\rm~1.75~\mu_{B}/Mn$ and $\rm~1.4~\mu_{B}/Mn$ for
$n=4$ and $n=5$ samples, respectively. This behavior is consistent
with double-exchange ferromagnetism as observed in previous
studies~\cite{AKZ2007,KLFI2002}. Because both of the constituents
LMO and SMO are insulating and antiferromagnetic (A type and G type
respectively) it seems likely, though this has not been proven, that
magnetism originates at the interfaces.

In Fig.~\ref{fig:Figure2} we show hard x-ray diffraction data for the $n=4$
superlattice. Thickness oscillations are visible, as well as several
superlattice reflections residing at integer values of the Miller
index $L$.  A key observation is that the $L=3$ reflection is
suppressed.  This is expected by symmetry.  For a superlattice with
sublayer thicknesses $m$ and $n$ the structure factor will vanish
for any reflection $L=u+v$ where $u/v$ is the fully reduced fraction
of $m/n$.  For the specific case of the $n=4$ superlattice the
structure factor for the $L=3$ reflection is:

\begin{eqnarray}
S(L=3)= 2 f_{int} - f_{LMO}-f_{SMO}\label{eqn:Equation1},
\end{eqnarray}
where $f_{LMO}$ and $f_{SMO}$ are the scattering factors of the
$\rm~MnO_{2}$ planes of the LMO and SMO layers, respectively, and
$f_{int}$ is the scattering factor for the $\rm~MnO_{2}$ plane at
the interface (Fig.~\ref{fig:Figure2})~\cite{TENSOR}. Notice that
the form factors for the $\rm~LaO$ and $\rm~SrO$ layers do not enter
this quantity. The $L=3$ reflection is forbidden by symmetry as long
as the interface $\rm MnO_{2}$ form factor is the average of the
form factors of $\rm~MnO_{2}$ planes of the LMO and SMO layers. The
intensity of this reflection is therefore a measure of the degree to
which the mirror symmetry of the interface is broken. That this
reflection is observed to be weak in hard x-ray diffraction
measurements is an indication that the interfaces are sharp, with
the symmetric form of Eq.~\ref{eqn:Equation1} not disrupted by
interdiffusion or other interface reconstruction.

Results of resonant scattering studies, which probe the unoccupied
density of states, reveal something surprising
(Fig.~\ref{fig:Figure2}a). If the photon energy is tuned above the O
K threshold - to a nonresonant condition - an $L$ scan reveals that
the $L=3$ is extinguished, in agreement with hard x-ray measurements.
However if the energy is tuned at $\rm530.2~eV$, near the Fermi
onset, the $L=3$ becomes visible. Evidently the symmetry of the
LMO-SMO interface, while preserved by the atomic lattice, is broken
electronically near the Fermi energy. This is evidence that the
interfaces are electronically reconstructed.

Our primary observation is the energy dependence of the $L=3$
reflection near the O K edge, shown in Fig.~\ref{fig:Figure3}. This
figure, which compares the intensity of the $L=3$ reflection to an
XAS spectrum of the LMO top layer, shows which part of the
unoccupied density of states is reconstructed. The $L=3$ reflection
has maximum intensity at the edge onset, where RSXS probes states at
the Fermi level. In addition, several weaker peaks are visible at
$531.5~\rm eV$ and $534~\rm eV$ which coincide roughly with the Mn
$e_{g,\uparrow}$ and $e_{g,\downarrow}$ bulk bands of LMO. We
conclude that electronic reconstruction of the LMO-SMO interface
occurs primarily near $E_F$ although higher-energy states might also
participate.

We propose that the main resonance observed at $\rm 529~eV$
corresponds to the presence of a quasiparticle peak at $E_{F}$, as
predicted for a Mott-band insulator interface~\cite{OM2004}. A
connection to the DMFT calculations for LTO-STO superlattices in
Ref.~\cite{OM2004} can be made by noting that, in the absence of
non-resonant scattering and excitonic effects, Im[$f_{int}$] is
proportional to the interface spectral function
$A_{int}(z=z_{int},z=z_{int};\omega)$. It was shown in
Ref.~\cite{OM2004} that $A_{int}(z=z_{int},z=z_{int};\omega)$ has a
strong peak at the Fermi energy $E_{F}$ as we observe here. A
quantitative comparison including the features at higher energy
would require a microscopic model of the superlattice combined with
Kramers-Kronig analysis of the data. The features at higher energy
might arise from mixing of high-energy degrees of freedom including
crystal field and Jahn-Teller effects, the Hubbard interaction and
the Hund's rule exchange interaction, which require microscopic
modeling beyond the scope of this article.

The connection between the $L=3$ resonance at $E_F$ and metallic
behavior is also supported by its temperature dependence, shown in
Fig.~\ref{fig:Figure1}b. As the temperature is lowered the intensity
of the $E_{F}$ resonance rises, the inflection point coinciding with
the peak in the resistivity and the onset of an in-plane magnetic
moment~\cite{AKZ2007} (Fig.~\ref{fig:Figure1}a). This is evidence
that the resonance is closely related to metallic behavior and that
the magnetism and metallic conduction both arise from reconstruction
of interfaces as considered by Okamoto \emph{et
al.}~\cite{OM2004,OM2005}. Interestingly, the $E_{F}$ peak also
appears to be composed of two features separated by $\rm 0.75~eV$
with both peaks following a similar temperature dependence.

In order to study the magnetism of the superlattice, we also
measured the $L=3$ reflection at the $\rm Mn~L_{3,2}$ edges.
Previous RSXS
studies~\cite{THGK2004,DMDO2004,WSBB2006,SSMK2005,OHMC2007} of
manganite systems have shown that both orbital and magnetic order
can contribute to scattering at these edges. In our structures, in
addition to magnetic order, one expects an interesting participation
of the $e_{g}$ orbital degree of freedom. In the SMO layer, which is
under $2.8 \%$ tensile strain, the $d_{x^2-y^2}$ orbital should be
lower in energy than the $d_{3z^2-r^2}$ orbital ($z$ is along the
c-axis). In LMO, which is under $2.1 \%$ compressive strain, the
situation is reversed, which favors an occupied $d_{3z^2-r^2}$
orbital. One therefore expects the total structure to have a
modulation in the orbital degree of freedom with the period of the
superlattice. This is in addition to the Jahn-Teller effect that may
still play a role in the LMO part of the superlattice despite the
presence of strain.~\cite{WBNM2006,AM2001} Separating charge,
orbital and magnetic scattering at the $\rm Mn~L$ edge through line
shape analysis can be a cumbersome task. However, this can be done
using the azimuthal dependence of the scattering. The magnetic
scattering is proportional to $(\hat{\epsilon}^{*}_{f} \times
\hat{\epsilon}_{i}) \cdot {\bf S}$, where $\hat{\epsilon}_{i}$ and
$\hat{\epsilon}_{f}$ are the incident and final polarizations,
respectively, and $\bf S$ is the local spin. For our experimental
geometry this simplifies to:

\begin{eqnarray}
I\propto\cos^2(\theta)\sin^2(\psi)+\sin^2(2\theta)\cos^2(\psi)\label{eqn:Equation2},
\end{eqnarray}
where $\theta$ is the angle of incidence on the sample. In contrast,
the strain-induced orbital scattering should be independent of
$\psi$ because strain does not break the tetragonal symmetry of the
unit cell. Therefore, to make a rough estimate of the relative size
of orbital and magnetic scattering, the $L=3$ reflection was
measured at a few values of the azimuthal angle $\psi$
(Fig.~\ref{fig:Figure4}). Good agreement is obtained between our
measurements at four $\psi$ angles and Eq.~\ref{eqn:Equation2}
(Fig.~\ref{fig:Figure4}b, inset) with a magnetic moment at $\sim
50^{\circ}$ between the a and b axes and a constant offset equal to
the off-resonance background. This suggests that the scattering is
primarily magnetic in the present case.

Finally we discuss the temperature dependence at the $\rm Mn~L$ edge
(Fig.~\ref{fig:Figure1}b), which agrees well with that observed at
the O K edge. This again suggests that the $E_{F}$ resonance and the
interfacial superlattice magnetization are interrelated. The overall
picture that emerges is that, as the temperature is lowered,
ferromagnetism nucleates at interfaces increasing the metallic
behavior and the spectral weight at $E_{F}$. The interfaces exhibit
the same connection between charge carrier itineracy and magnetic
order familiar from bulk colossal magnetoresistance materials.

In conclusion, we have used RSXS to investigate superlattices of the
Mott insulator $\rm LaMnO_{3}$ and the ``band" insulator $\rm
SrMnO_{3}$. By choosing a specific combination of superlattice
reflection and layer periods we were able to isolate the electronic
properties of the interface. We show that the interface density of
states exhibits a pronounced peak at $E_{F}$ as shown in the
calculations of Okamoto \emph{et al.} suggesting that ferromagnetism
and metallic behavior in this system arise from electronic
reconstruction of the interfaces. Our study demonstrates a general
strategy for using RSXS to probe the electronic properties of
interfaces.

The authors acknowledge helpful discussions with John Freeland and
C.-C. Kao. This work was supported by the Office of Basic Energy
Sciences at the U.S. Department of Energy. Resonant scattering
experiments were supported under grant No. DE-FG02-06ER46285 and
digital synthesis work under grant No. DE-AC02-06CH11357 subcontract
WO 4J-00181-0004A. Work at Argonne was supported under contract No.
DE-AC02-06CH11357 and use of the NSLS under contract No.
DE-AC02-98CH10886.


\begin{thebibliography}{}
\bibitem{TENSOR}
The atomic scattering factors are tensors, i.e. $f=f^{i,j}$. For
notational simplicity, however, we leave off the indices in our
discussion.
\end{thebibliography}
\end{document}